\documentclass[english,11pt]{article}
\usepackage{epsf,amsmath,amssymb,graphicx,dcolumn}
\usepackage{scalefnt,subfigure,ulem,pstricks}
\usepackage{cite,hyperref}
\usepackage{todonotes}
\usepackage{color}
\usepackage{rotating}
\definecolor{lightblue}{rgb}{.7,.8,1}
\newcommand{\pdf}{{\abbrev PDF}}
\newcommand{\qcd}{{\abbrev QCD}}

\newcommand{\abbrev}{\scalefont{.9}}
\newcommand{\pt}[1]{p_\text{T#1}}

\newcommand{\muF}{\mu_\text{F}}
\newcommand{\muR}{\mu_\text{R}}
\newcommand{\mhiggs}{m_H}
\newcommand{\mtop}{m_t}

\newcommand{\eqn}[1]{Eq.\,(\ref{#1})}
\newcommand{\fig}[1]{Fig.\,\ref{#1}}

\newcommand{\dd}{{\rm d}}

\newcommand{\order}[1]{{\cal O}(#1)}

\newcommand{\lo}{\text{\abbrev LO}}
\newcommand{\nlo}{\text{\abbrev NLO}}
\newcommand{\nnlo}{\text{\abbrev NNLO}}

\newcommand{\msbar}{\overline{\mbox{\abbrev MS}}}

\title{\vspace*{-6em}
  \begin{flushright}
    {\sf\small
      CERN-PH-TH/2012-146 
      --- LPN12-053 
      --- UCLA/12/TEP/101
      --- WUB/12-13 
    }
  \end{flushright}
\vspace*{2em} Top-mass effects in differential Higgs production
  through gluon fusion at $\order{\alpha_s^4}$}
\author{Robert V. Harlander$^a$, Tobias Neumann$^a$, Kemal
  J. Ozeren$^b$,\\ Marius Wiesemann$^{a,c}$\\[2em]
$^a$ {\it Fachbereich C,
  Bergische Universit\"at Wuppertal,}\\[-.3em] {\it 42097 Wuppertal,
  Germany}\\
$^b$ {\it Department of Physics and Astronomy, UCLA,}\\[-.3em]
{\it Los Angeles, CA 90095-1547, USA}\\
$^c$ {\it TH Division, Physics Department, CERN}\\[-.3em]
{\it CH-1211 Geneva 23, Switzerland}\\
{\small\tt harlander@physik.uni-wuppertal.de}\\[-.3em]
{\small\tt tobias.neumann@uni-wuppertal.de}\\[-.3em]
{\small\tt marius.wiesemann@cern.ch}\\[-.3em]
 {\small\tt ozeren@physics.ucla.edu }
}

\date{}
\begin{document}
\maketitle

\begin{abstract}
Effects from a finite top quark mass on differential distributions in
the Higgs+jet production cross section through gluon fusion are studied
at next-to-leading order in the strong coupling,
i.e.\ $\order{\alpha_s^4}$. Terms formally subleading in $1/\mtop{}$ are
calculated, and their influence on the transverse momentum and rapidity
distribution of the Higgs boson are evaluated. We find that, for the
differential K-factor, the heavy-top limit is valid at the 2-3\% level
as long as the transverse momentum of the Higgs remains below about
150\,GeV.
\end{abstract}

\section{Introduction}

The past few years in particle physics have been characterized by ever
more sensitive exclusion limits for Higgs bosons (see, e.g.,
Refs.\cite{TEVNPH:2012ab,ATLAS:2012ae,Chatrchyan:2012tx}). These
results were based on the combination of experimental data, background
calculations and extrapolations, theoretical expectations for the
signal, and careful estimates of the associated
uncertainties\cite{Dittmaier:2011ti,Dittmaier:2012vm}. Concerning the
signal cross sections, these uncertainties have several sources: the
parton densities (\pdf{}s), the strong coupling $\alpha_s(M_Z)$, and
higher order perturbative effects, for example.

An uncertainty which is very specific to hadronic Higgs production in
the Standard Model concerns the error induced by evaluating higher order
perturbative corrections in an effective theory, derived by letting
$\mtop\to\infty$, where $\mtop$ is the top quark mass. The observation
that the next-to-leading order (\nlo{}) corrections to the inclusive
total cross section are approximated at the percent
level\cite{Spira:1995rr,Dawson:1993qf} in this approach has been used as
an argument for trusting it also at higher orders: next-to-\nlo{}
(\nnlo{}) \qcd{}
corrections\cite{Harlander:2002wh,Anastasiou:2002yz,Ravindran:2003um}
have thus led to a perturbatively robust prediction for this
quantity.\footnote{Effects beyond \nnlo{} and electro-weak corrections
  have been studied in
  Refs.\cite{Catani:2003zt,Idilbi:2005ni,Ravindran:2006cg,Ahrens:2008nc,
    Djouadi:1994ge,Degrassi:2004mx,Aglietti:2004nj,Actis:2008ug,
    Anastasiou:2008tj}, for example.} A few years ago, the effective
theory approach was tested at \nnlo{} by an explicit calculation of the
subleading terms in $1/\mtop{}$ to the total inclusive cross
section\cite{Harlander:2009mq,Harlander:2010my,Pak:2009dg,Pak:2011hs}. It
was found that these terms have an effect of less than 1\%.

However, this result does not allow for a direct generalization to less
inclusive quantities. Since they depend on several kinematical
parameters such as the transverse momentum $\pt{}$, the rapidity $y$, or
simply phase space cuts, such observables may have a very different
convergence behavior in $1/\mtop{}$ than the inclusive cross section.

Nevertheless, so far the \nlo{} $\pt{}$- and $y$-distributions in
$H+$jet-production\cite{deFlorian:1999zd,Glosser:2002gm,%
  Ravindran:2002dc,deFlorian:2005rr}, the jet-vetoed Higgs cross
section\cite{Catani:2001cr}, as well as the resummation of the
logarithmic terms for small
$\pt{}$\cite{Bozzi:2003jy,Bozzi:2005wk,Bozzi:2007pn,Catani:2010pd} are
based on the effective theory approach, of course, and so are the fully
exclusive \nnlo{} partonic Monte Carlo programs for Higgs production in
gluon fusion\cite{Anastasiou:2004xq,Catani:2008me,Catani:2007vq}.

Only rather few studies have been aimed at quantitatively testing or
going beyond the heavy-top approximation in Higgs distributions. Such
results are only available at leading order (\lo{}) in perturbation
theory, for $H+n$jet with
$n=0,1,2$\cite{DelDuca:2003ba,DelDuca:2001fn,DelDuca:2001eu,Alwall:2011cy,Bagnaschi:2011tu}.
Generally speaking, one finds that the approximation works rather well
for $\pt{}<\mtop$.

In this paper, we evaluate the $1/\mtop{}$-effects to $H+$jet production
at \nlo{} \qcd{}. We focus on the initial states $gg$ and $qg$ which, in
the $\msbar$ scheme, are typically about two orders of magnitude larger
than all other channels combined. We will show that the radiative
corrections to the mass effects in the $gg$ channel are remarkably close
to those of the heavy-top limit. In the sum over all partonic
sub-processes, this is deteriorated to some extent by the $qg$-channel
which is numerically subleading, however. As a result, we find that the
kinematical distributions can be calculated with 2-3\% accuracy by
reweighting the \lo{} distributions (including the full top mass
dependence) by the differential \nlo{} K-factor evaluated in the
heavy-top mass limit.

The remainder of the paper is organized as follows: In
Section~\ref{sec::outline}, we briefly specify the problem under
consideration by introducing the relevant Feynman diagrams;
Section~\ref{sec::lo} presents some \lo{} motivation of our study;
Section~\ref{sec::nlo} contains the main part of the paper, including
the results for the $\pt{}$-integrated cross section as well as for the
$\pt{}$ and rapidity distributions, all at \nlo{}; our conclusions are given
in Section~\ref{sec::conclusions}.

\section{Outline of the problem}\label{sec::outline}

\begin{figure}
  \begin{center}
    \begin{tabular}{ccc}
      \mbox{\includegraphics[height=.12\textheight]{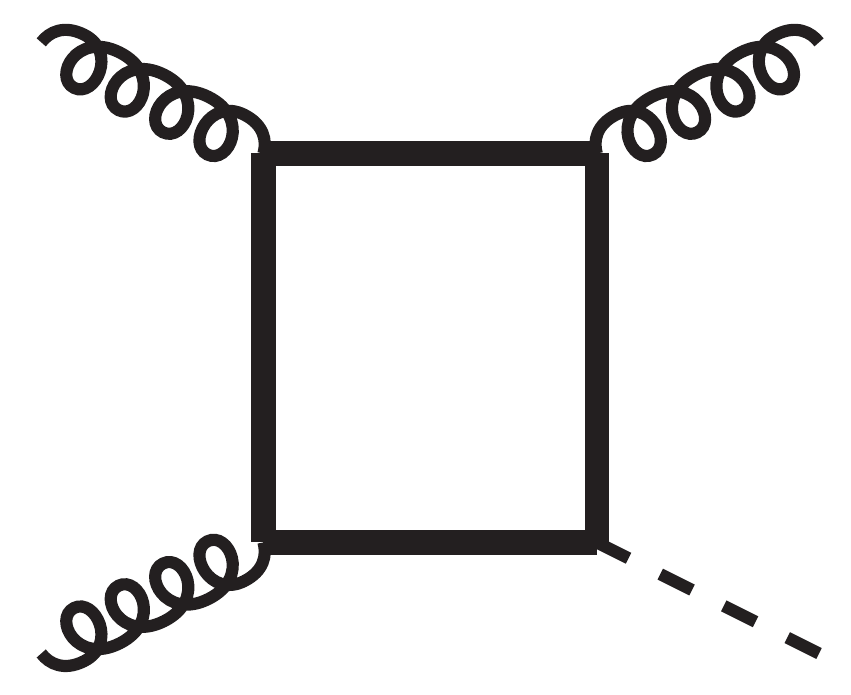}} &
      \mbox{\includegraphics[height=.12\textheight]{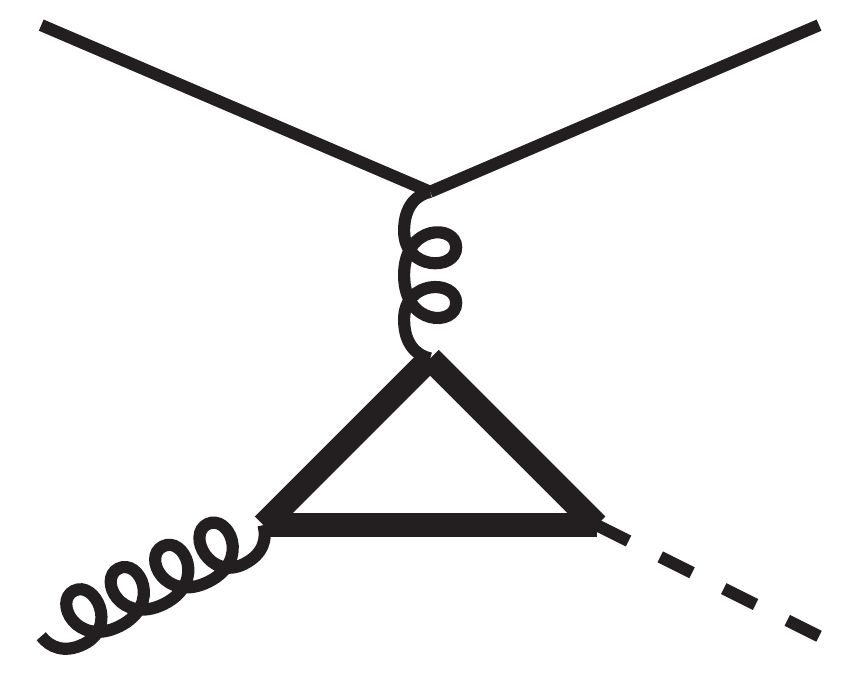}} &
      \raisebox{.7em}{\includegraphics[height=.12\textheight]{%
          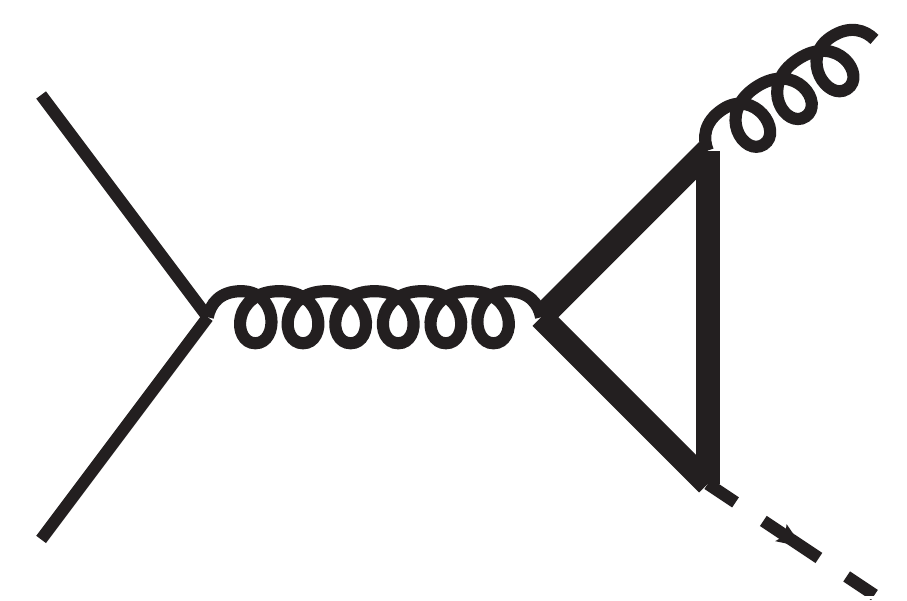}}
      \\
      (a) & (b) & (c)
    \end{tabular}
    \parbox{.9\textwidth}{
      \caption[]{\label{fig::diaslo}\sloppy Sample Feynman diagrams
        contributing to the process $pp\to H+$jet at \lo{} \qcd{}. The
        graphical notation for the lines is: thick straight $\hat=$ top
        quark; thin straight $\hat=$ light quark $q\in\{u,d,c,s,b\}$;
        spiraled $\hat=$ gluon; dashed $\hat=$ Higgs boson.
} }
  \end{center}
\end{figure}

In this paper we consider the quantities $\dd\sigma/\dd \pt{}$ and
$\dd\sigma/\dd y$ in the gluon fusion process, where a Higgs boson is
produced in association with a jet in hadronic collisions through a
top-loop mediated gluon-Higgs coupling. Other quark-loop contributions
are suppressed by their Yukawa coupling and will be neglected. The
Higgs' transverse momentum $\pt{}$ and its rapidity $y$ are measured
relative to the hadronic center-of-mass system. The \lo{} contribution
to this process is of order $\alpha_s^3$; it is obtained by convolving
the partonic subprocesses $gg\to Hg$, $qg\to Hq$, $\bar qg\to H+\bar q$,
and $q\bar q\to Hg$ ($q\in\{u,d,s,c,b\}$), see \fig{fig::diaslo}, with
the corresponding parton density functions. At this order of
perturbation theory, the full dependence on the top quark and Higgs
boson mass $\mhiggs$ is known, and also parton shower effects have been
evaluated\cite{Bagnaschi:2011tu,Alwall:2011cy}.

\begin{figure}
  \begin{center}
    \begin{tabular}{cccc}
      \mbox{\includegraphics[height=.12\textheight]{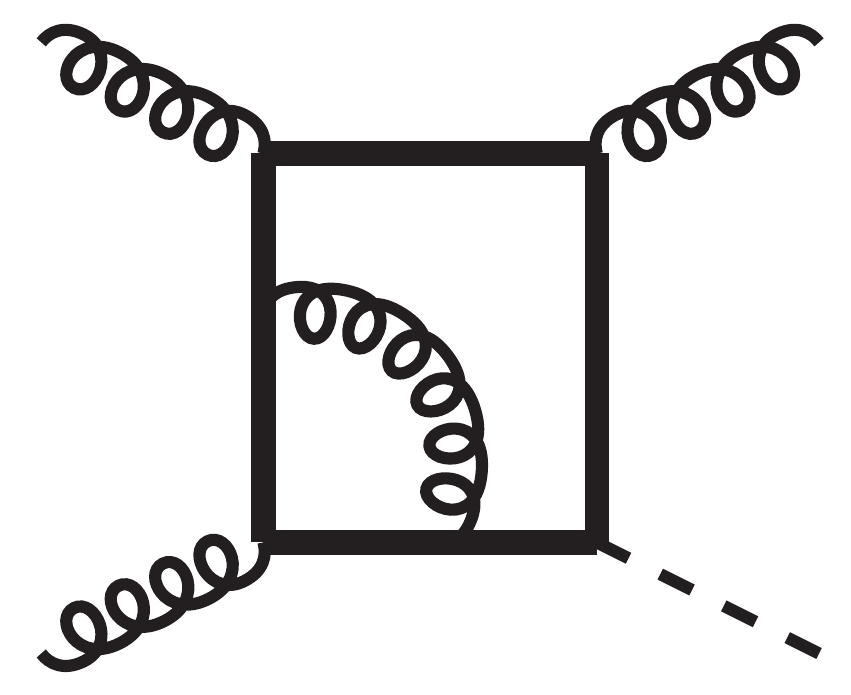}} &
      \mbox{\includegraphics[height=.12\textheight]{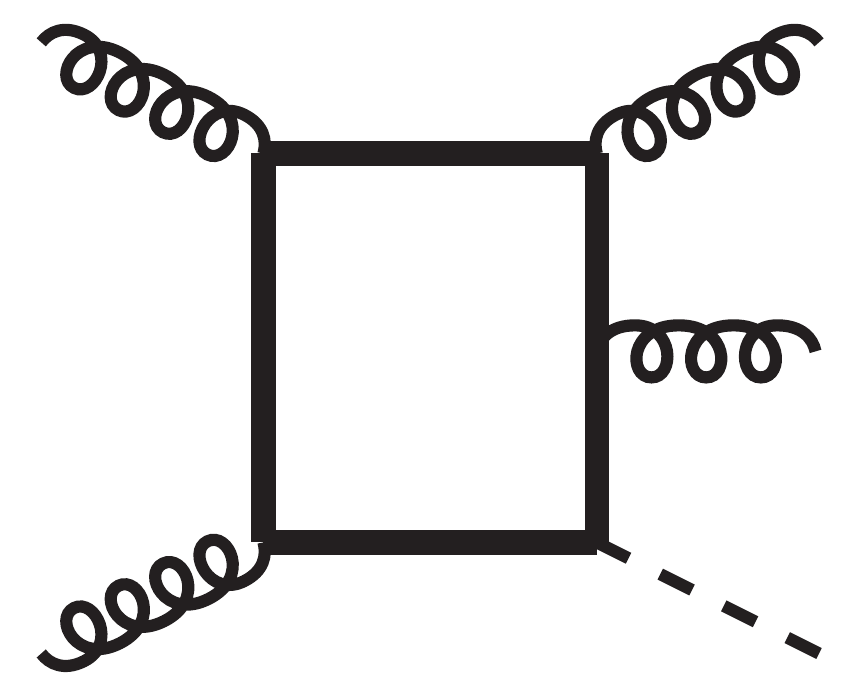}} &
      \mbox{\includegraphics[height=.12\textheight]{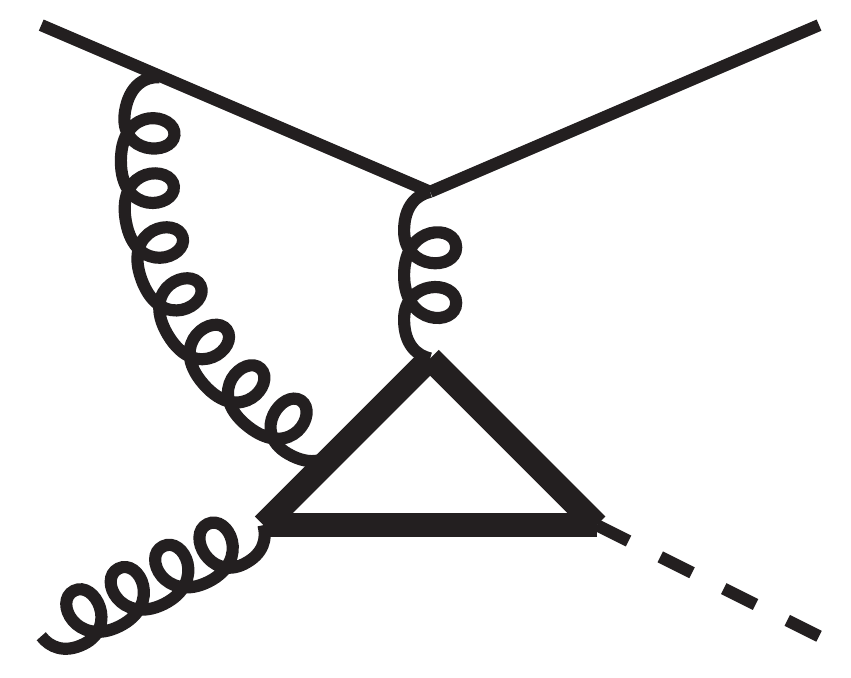}} &
      \mbox{\includegraphics[height=.12\textheight]{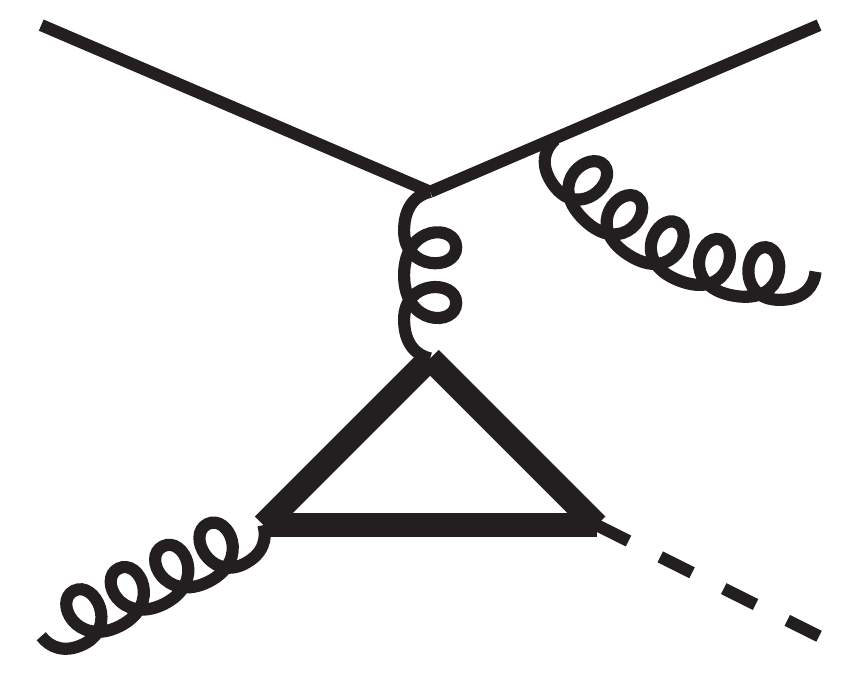}}
      \\
      (a) & (b) & (c) & (d)
    \end{tabular}
    \begin{tabular}{cc}
      \mbox{\includegraphics[height=.12\textheight]{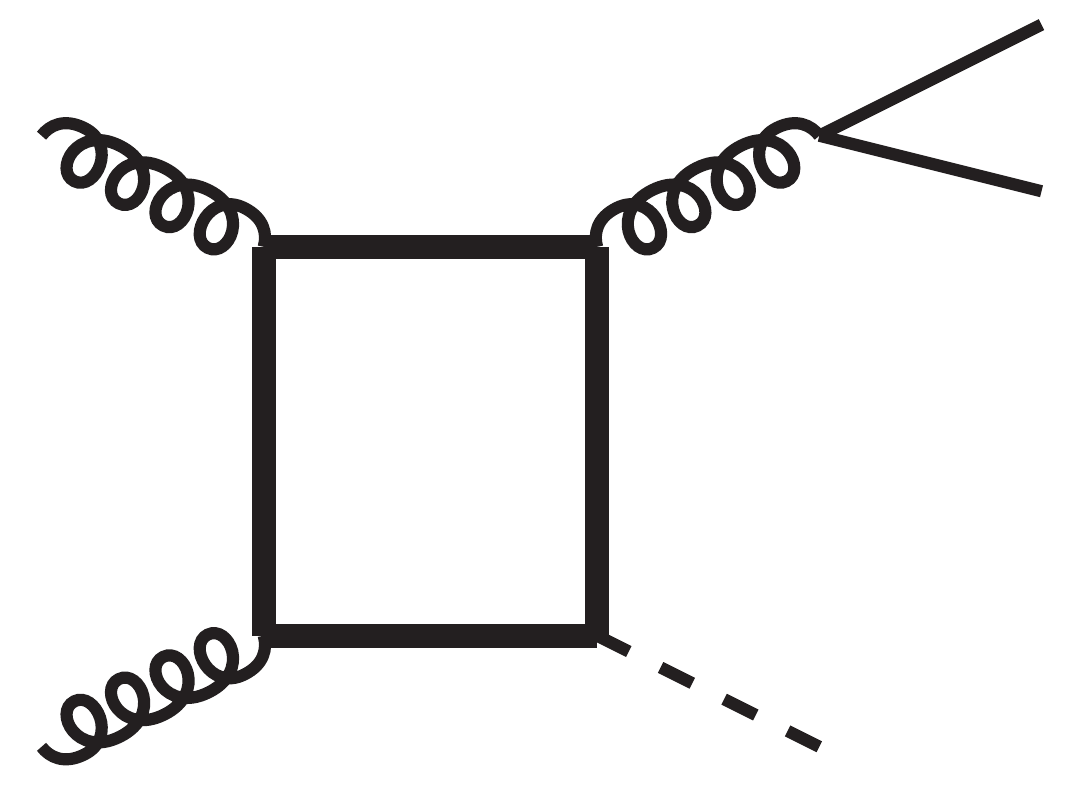}} &
      \raisebox{.2em}{\includegraphics[height=.12\textheight]{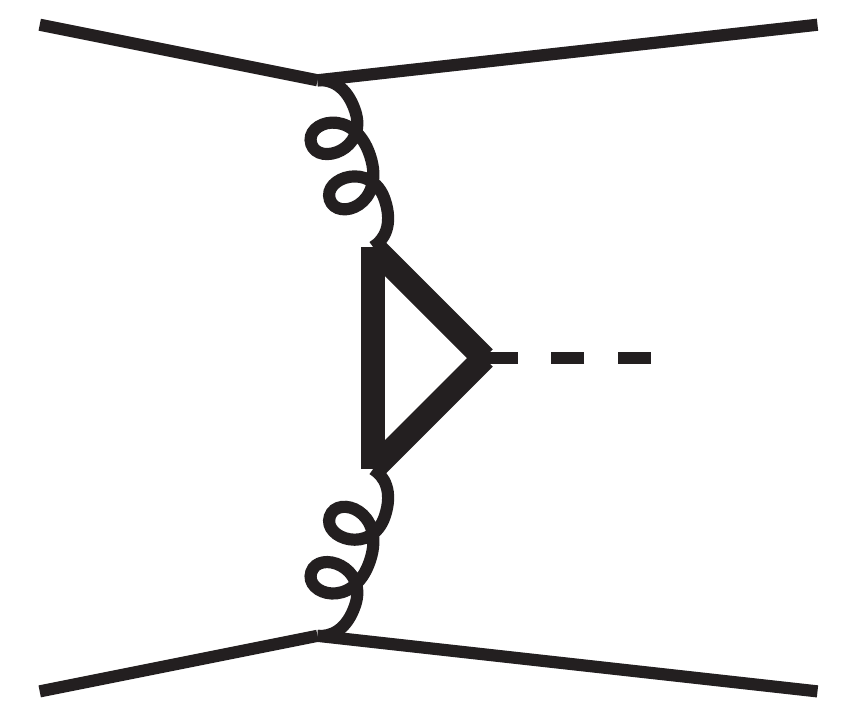}} 
      \\
      (e) & (f)
    \end{tabular}
    \parbox{.9\textwidth}{%
      \caption[]{\label{fig::diasnlo}\sloppy
        Sample Feynman diagrams contributing to the
      process $pp\to H+$jet at \nlo{} \qcd{}. Notation as in
      \fig{fig::diaslo}.}
      }
  \end{center}
\end{figure}

\begin{figure}
  \begin{center}
    \begin{tabular}{c}
      \mbox{\includegraphics[height=.4\textheight]{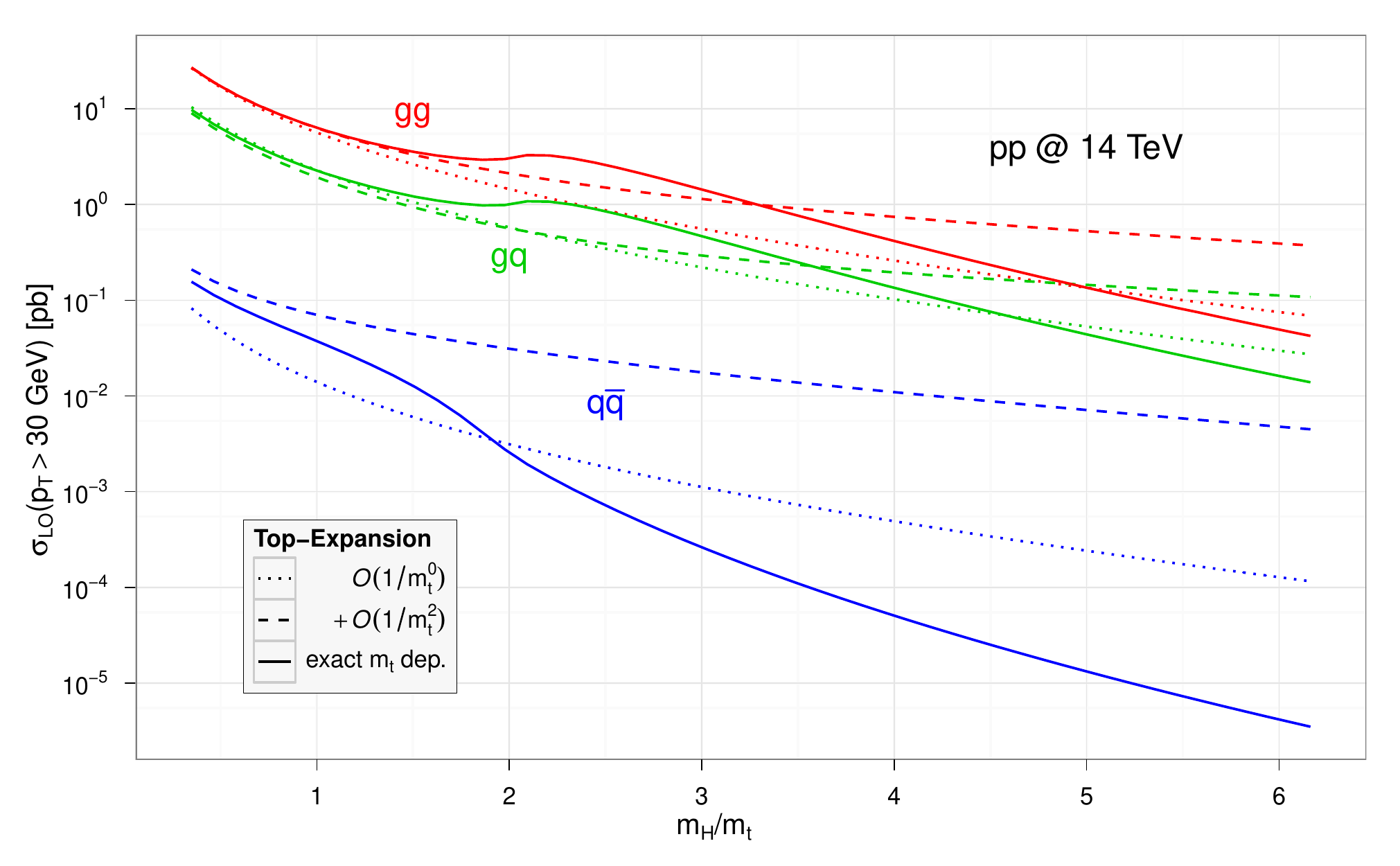}}
    \end{tabular}
    \parbox{.9\textwidth}{
      \caption[]{\label{fig::sigmh}\sloppy Higgs+jet cross section as
        defined in \eqn{eq::dsdpt}, with $\pt{}^\text{cut} = 30$\,GeV.}
      }
  \end{center}
\end{figure}

At \nlo{}, the Feynman diagrams can be divided into three groups: the
first one is obtained by dressing each of the partonic \lo{} processes
by a virtual or a real gluon, see \fig{fig::diasnlo}\,(a)-(d), for
example; the second one by splitting the emitted gluon into a $q\bar
q$-pair, see \fig{fig::diasnlo}\,(e). The third group is of the form
$q_1q_2\to Hq_1q_2$, where both $q_1$ and $q_2$ run continuously from
the inital to the final state, and $q_1,q_2$ denote quarks or
anti-quarks of the first five flavors, see \fig{fig::diasnlo}\,(f), for
example.

\section{Leading order considerations}\label{sec::lo}

\fig{fig::sigmh} shows the \lo{} result for the cross section
\begin{equation}
\begin{split}
\sigma(\pt{}>\pt{}^\text{cut})=\int_{\pt{}\geq\pt{}^\text{cut}}\dd
\pt{}\,\frac{\dd\sigma}{\dd\pt{}}
\label{eq::dsdpt}
\end{split}
\end{equation}
as a function of $\mhiggs$, divided into the individual partonic
sub-processes according to the $\msbar$-scheme, and keeping the full top
mass dependence (solid), the expansion in $1/\mtop$ through $1/\mtop^0$
(dotted), and through $1/\mtop^2$ (dashed).  Unless stated otherwise, we
will set $\pt{}^\text{cut}=30$\,GeV in this paper; also, we choose the
renormalization and factorization scales to be $\muR=\muF=\mhiggs$; the
on-shell top quark mass is set to $\mtop=172$\,GeV, and the default
hadronic center-of-mass energy is $\sqrt{s}=14$\,TeV, but we will
include exemplary results for $\sqrt{s}=7$\,TeV below.

The kink in the cross section at $\mhiggs\approx 2\mtop$ in
\fig{fig::sigmh} is due to the top-quark threshold in the scattering
amplitude. Clearly, this structure cannot be reproduced by an expansion
in $1/\mtop$. Note that the shape of the curve is very reminiscent of
the total inclusive cross section $pp\to H+X$ through gluon fusion
which, however, is a $2\to 1$ process at \lo{} and starts at
$\order{\alpha_s^2}$; the quantity displayed here is the
cross section for $H+$jet production and therefore of
$\order{\alpha_s^3}$.

As is obvious from \fig{fig::sigmh}, the $1/\mtop$-expansion for the
$q\bar q$ channel is significantly worse than for the other two
channels. This failure of the $1/\mtop$-expansion to reproduce the
$q\bar q$-channel has already been observed for the total inclusive
cross section in Refs.\cite{Harlander:2010my,Pak:2011hs}.  However, the
$q\bar q$-channel is about two orders of magnitude smaller than $qg$
which itself is a factor of 2-3 smaller than $gg$. Similar observations
hold for the other purely quark induced channels which enter at higher
orders, specifically $qq$, $qq'$, and $q\bar q'$. The conclusion to draw
from this is that the quark-induced channels constitute a solid, but
rather minor limitation of the heavy-top limit. Our analysis cannot
bring any further insights for these channels, and we will disregard
them in what follows. One should keep in mind, however, that kinematical
cuts could enhance the pure quark channels; in this case, results based
on the heavy-top limit become unreliable.

\begin{figure}
  \begin{center}
    \begin{tabular}{c}
      \mbox{\includegraphics[height=.4\textheight]{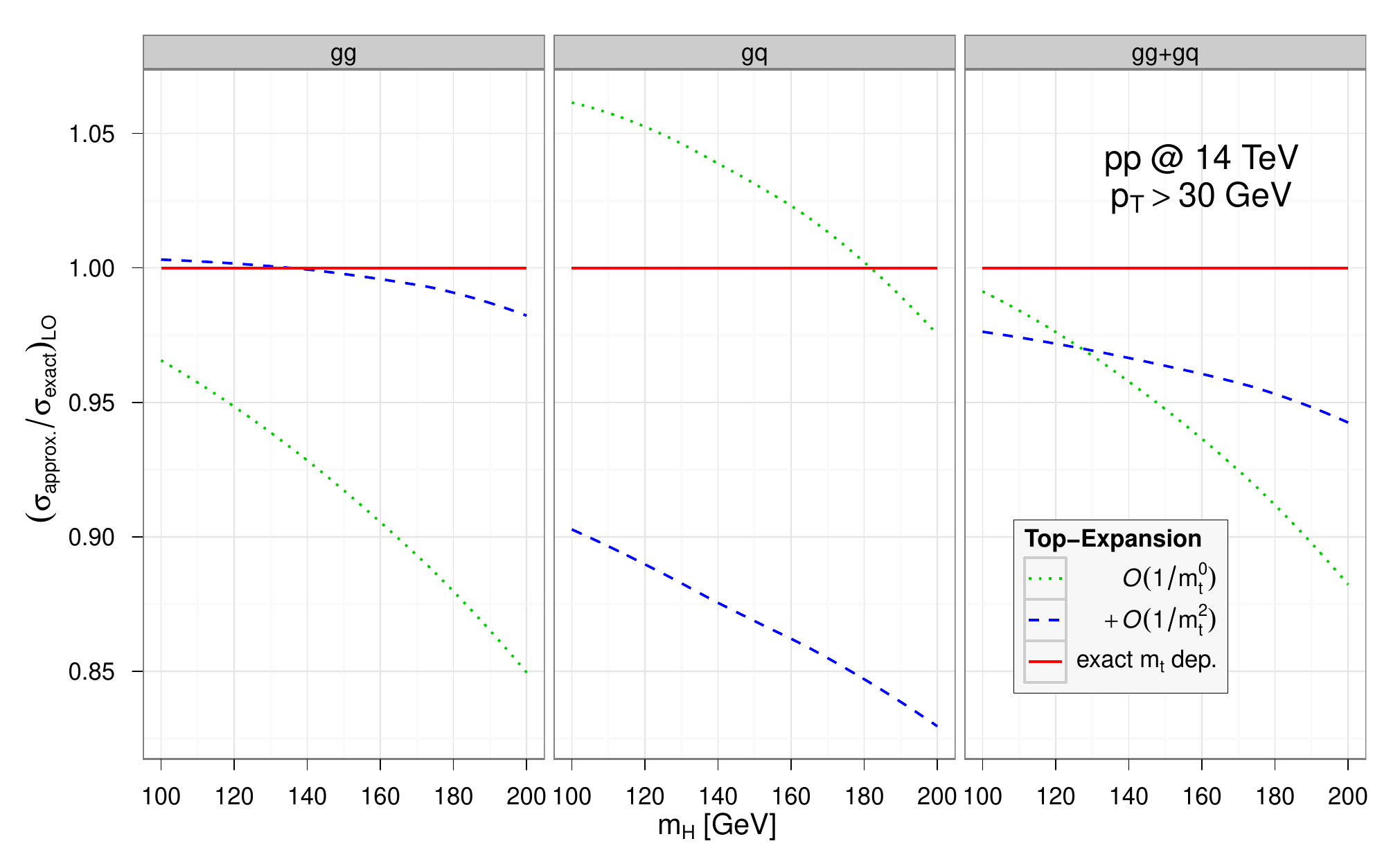}}\\[-.5em]
      (a)\\
      \mbox{\includegraphics[height=.4\textheight]{%
          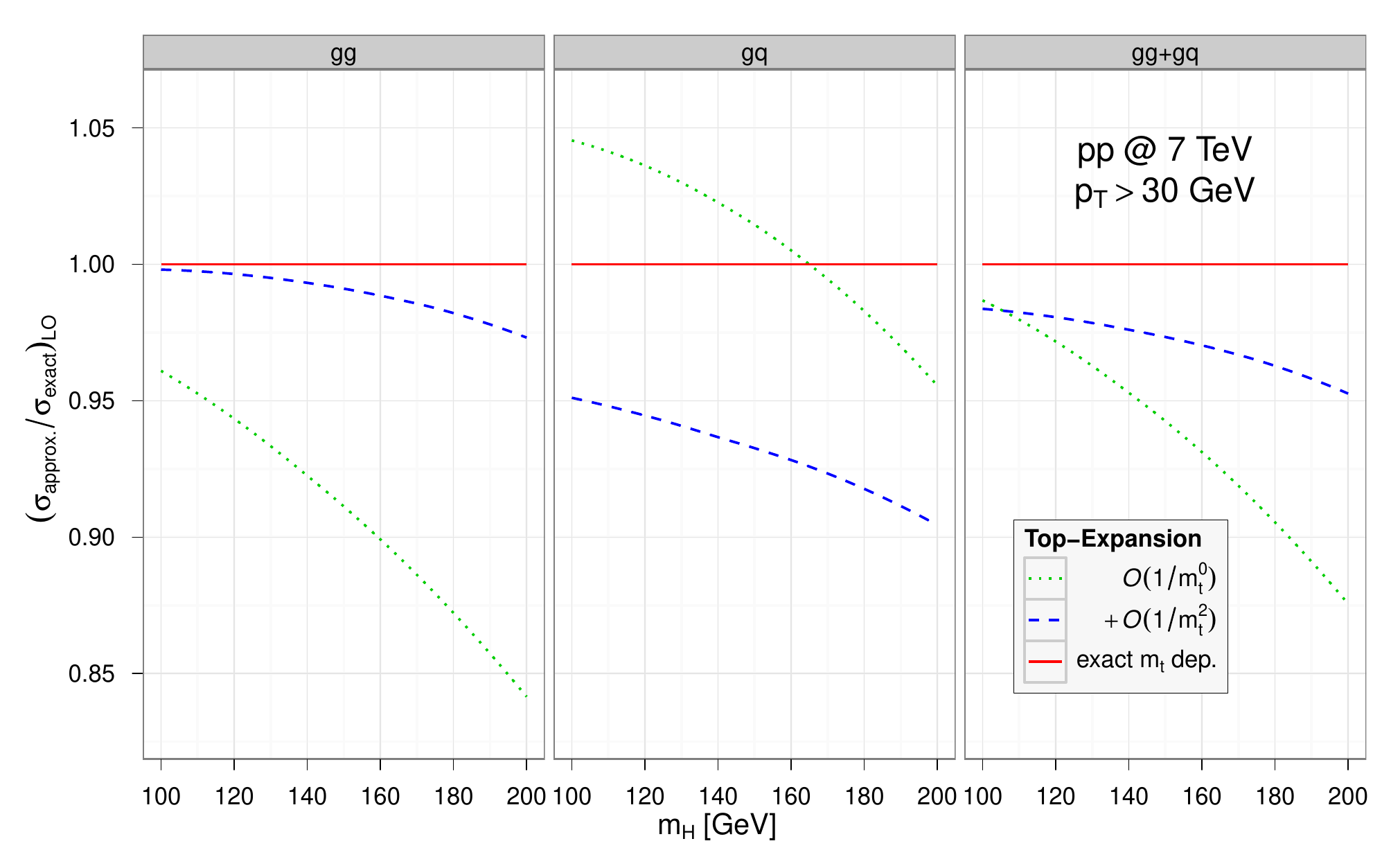}}\\[-.5em]
      (b)
    \end{tabular}
    \parbox{.9\textwidth}{
      \caption[]{\label{fig::LmH}\sloppy Ratio of the integrated \lo{}
        cross section from \eqn{eq::dsdpt} when expanded through
        $1/\mtop^n$ to the exact result, for $n=0$ (dotted) and $n=2$
        (dashed). Left: only $gg$; center: only $qg$; right: sum of $gg$
        and $qg$; (a) $\sqrt{s}=14$\,TeV --- (b) $\sqrt{s}=7$\,TeV.}}
  \end{center}
\end{figure}

The relative deviation of the $1/\mtop{}$ expansion from the exact
result (still at \lo{}) is shown in \fig{fig::LmH}\,(a). The curves are
obtained by dividing the integrated cross section as defined in
\eqn{eq::dsdpt} when expanded in $1/\mtop{}$ by the same expression when
the full top mass dependence is kept. The individual plots show this
ratio separately for the case when only the $gg$-channel (left) and only
the $qg$-channel (center) is taken into account (both in the numerator
and the denominator of the ratio), and also for the sum of both channels
(right). At this point, despite our default choice for the
center-of-mass energy of $\sqrt{s}=14$\,TeV, we also include results for
$\sqrt{s}=7$\,TeV in order to obtain an impression of the dependence of
our results on $\sqrt{s}$. The corresponding plots are shown in
\fig{fig::LmH}\,(b). In the following discussion, the numbers for
$\sqrt{s}=7$\,TeV are referred to in brackets; the unbracketed numbers
are for the default $\sqrt{s}=14$\,TeV.

For the $gg$-channel, the heavy-top expansion through
$\order{1/\mtop^2}$ approximates the exact result up to 2\% (2.5\% for
$\sqrt{s}=7$\,TeV) within $\mhiggs\in[100,200]$\,GeV, while the leading
$\order{1/\mtop^0}$ term deviates up to 15\% (16\%) from it.  For the
$qg$-channel, the approximation by the $1/\mtop^2$-result is not as
good: while the $1/\mtop^0$ term remains within about 7\% (5\%) of the
exact result, the $1/\mtop^2$-term deviates by up to 18\%
(9.5\%). However, since the $gg$-channel is numerically dominant, for
the sum of both channels the $1/\mtop^2$-approximation agrees with the
full result to better than 6\% (5\%), while the difference between the
$1/\mtop^0$ and the exact result ranges up to 12\% (12.5\%).

We see that the dependence of the $1/\mtop$-effects on the
center-of-mass energy is very weak for the dominant $gg$-channel; for
the $qg$-channel, the $1/\mtop$-effects are more pronounced for higher
$\sqrt{s}$ due to the fact that, technically, $\mtop$ is always assumed
to be the largest scale in the problem (for a more detailed discussion,
see
Ref.\,\cite{Harlander:2009bw,Pak:2009bx,Harlander:2009mq,Pak:2009dg}). In
order to study the quality of the heavy-top approximation, it is
therefore sufficient to use our default setting $\sqrt{s}=14$\,TeV.

The absolute size of the mass effects at \lo{} is used here only as an
indicator of how far we can expect to be able to trust the heavy-top
expansion at \nlo{}. This indicator can be considered as a lower limit
for the validity range though: In the total inclusive cross section, it
has proved useful to factor out the \lo{} top mass dependence from the
perturbative corrections. Our results will show that it is very
advantageous to follow this strategy also for differential cross
sections.

\begin{figure}
  \begin{center}
    \begin{tabular}{c}
      \mbox{\includegraphics[height=.4\textheight]{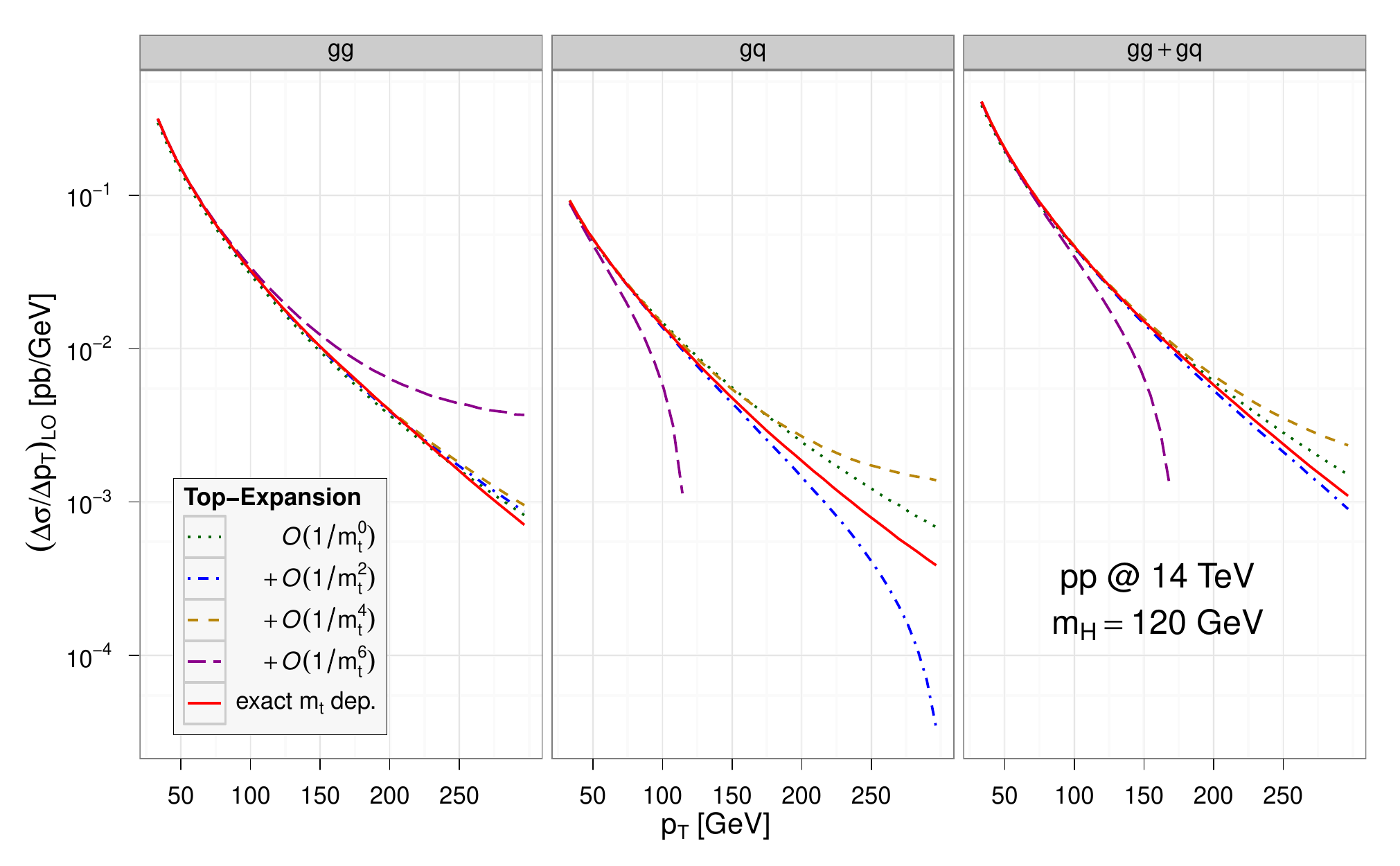}}
    \end{tabular}
    \parbox{.9\textwidth}{
      \caption[]{\label{fig::pTLO}\sloppy Differential cross section
        $\dd\sigma/\dd\pt{}$ at \lo{} \qcd{}, where $\pt{}$ is the
        transverse momentum of the Higgs boson. Solid curve: full
        $\mtop{}$-dependence included;
        dotted/dash-dotted/short-dashed/long-dashed: expansion in
        $1/\mtop^n$ with $n=0/2/4/6$. Left, center, and right plot
        show the $gg$-, the $qg$-channel, and their sum, respectively.}}
  \end{center}
\end{figure}

Turning to more exclusive quantities, \fig{fig::pTLO} compares the exact
result for the Higgs' transverse momentum distribution to expansions
including successively higher orders in $1/\mtop$, separately for the
$gg$- and the $qg$-channel. 
The $1/\mtop$-expansion works very well, roughly up to $\pt{}=\mtop$, as
long as one restricts oneself to lower orders in $1/\mtop$. At
$\order{1/\mtop^6}$ and beyond, convergence seems to be lost at much
lower values of $\pt{}$. This is due to the region of large partonic
center-of-mass energy $\sqrt{\hat s}$: similar to the calculation of
Ref.\cite{Harlander:2009mq,Harlander:2010my,Pak:2009dg,Pak:2011hs},
the $1/\mtop$ expansion generates a power behavior in $\hat
s/\mtop^2$. At lower orders in $1/\mtop$, these terms are suppressed by
the parton densities. At higher orders, however, they spoil the
convergence behavior of the hadronic cross section. In our \nlo{}
analysis, we therefore restrict ourselves to comparisons of the
$1/\mtop^0$ to the $1/\mtop^2$ terms.

\begin{figure}
  \begin{center}
    \begin{tabular}{c}
      \mbox{\includegraphics[height=.4\textheight]{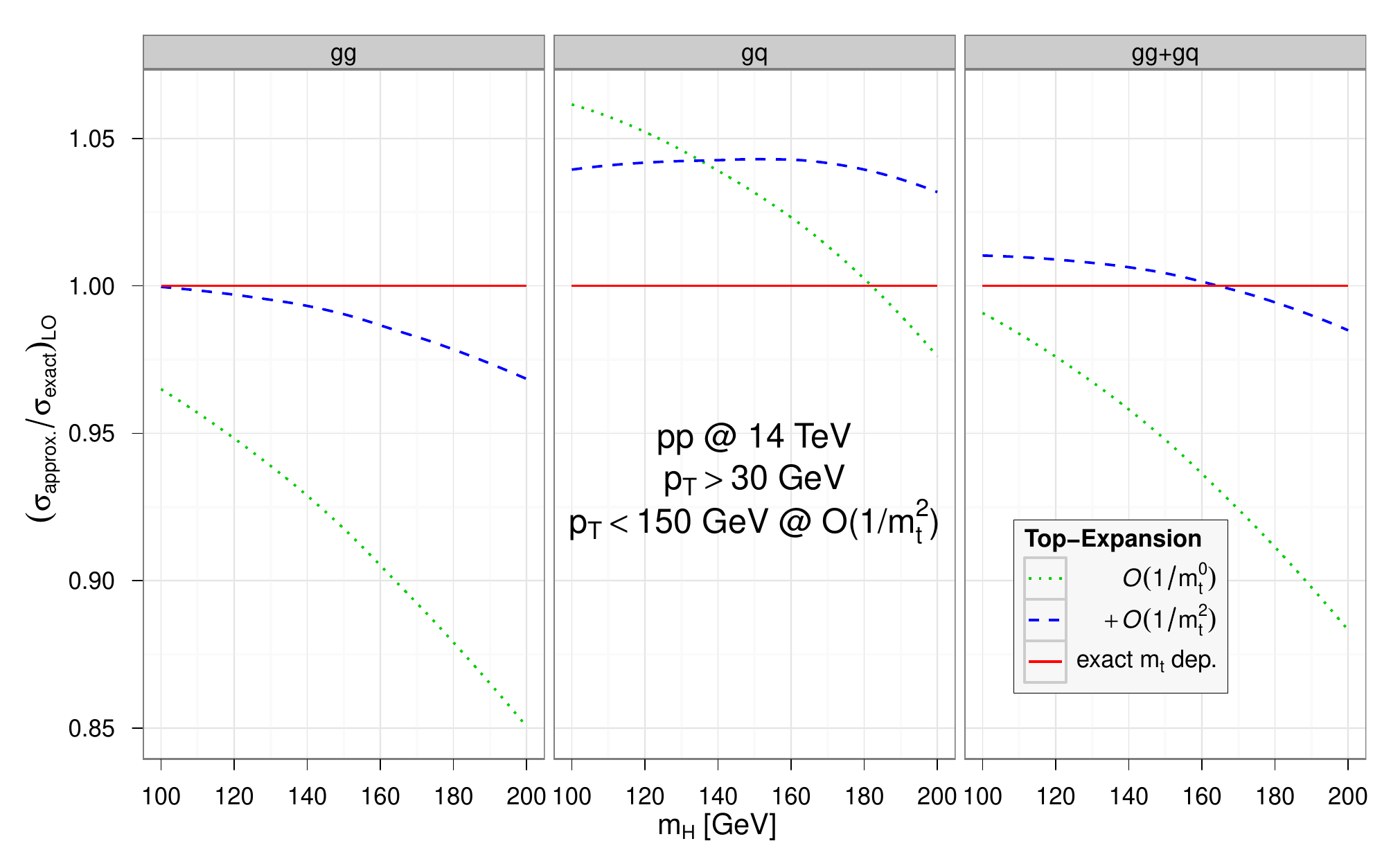}}
    \end{tabular}
    \parbox{.9\textwidth}{
      \caption[]{\label{fig::sigmhpt150}\sloppy Similar to
        \fig{fig::LmH}, but with an upper $\pt{}$-cut for the
        $1/\mtop^2$-coefficient: $1/\mtop^2\to
        1/\mtop^2\cdot\Theta(150-\pt/\text{GeV})$.}}
  \end{center}
\end{figure}

The behavior of $\dd\sigma/\dd\pt{}$ suggests to try to improve the
approximation of the integrated cross section
$\sigma(\pt{}>\pt{}^\text{cut})$ by discarding the $1/\mtop^2$-effects
above $\pt{}\gtrsim 150$\,GeV, and taking into account only the leading
$1/\mtop^0$ terms for larger values of $\pt{}$. The result is shown in
\fig{fig::sigmhpt150}. While the effect of this procedure is small in
the $gg$-channel, the contribution from $\pt{}>150$\,GeV is much more
significant in the $qg$-channel. Even though the numerical approximation
improves, in particular for the sum $gg+qg$, this rather large effect
indicates that the result depends quite strongly on the specific upper
cut on $\pt{}$ which has been introduced for the $1/\mtop^2$ terms. We
conclude that this does not allow for a systematic improvement and will
not consider it any further.

Overall, the \lo{} observations are encouraging to study the behavior of
the $1/\mtop$ terms at \nlo{} in order to estimate the validity range of
the heavy-top limit also for differential quantities.

\section{Next-to-leading order results}\label{sec::nlo}

\subsection{Outline of the calculation}

The most complicated Feynman diagrams are of the two-loop box-type with
massless and massive (mass $\mtop$) internal and one massive external
line (mass $\mhiggs$), see Figs.\,\ref{fig::diasnlo}\,(a) and (c), for
example. Although not out of reach, the complexity of the corresponding
integrals is too high for an efficient numerical evaluation. Therefore,
to date the \nlo{} corrections to this process are only available in an
effective theory approach where the top quark is integrated
out\cite{deFlorian:1999zd,Glosser:2002gm,%
  Ravindran:2002dc,deFlorian:2005rr}. The Feynman diagrams then simplify
to one-loop level, with an effective Higgs-gluon vertex, multiplied by a
Wilson coefficient which can be evaluated
perturbatively\cite{Chetyrkin:1997un,Kramer:1996iq,%
  Chetyrkin:1997iv,Schroder:2005hy,Chetyrkin:2005ia}.

The effective theory approach can be seen as the leading term of an
expansion for small $1/\mtop$. The goal of this paper is to go beyond
this limit and to study the behavior of the next term in this expansion.
In Ref.\cite{Harlander:2009mq}, the relevant one-loop $2\to 2$ and
tree-level $2\to 3$ amplitudes have been obtained through automated
asymptotic expansions\cite{Harlander:1997zb,%
  Steinhauser:2000ry,Smirnov:2002pj}. For our purposes, we combine them
here with the dipole subtraction terms\cite{Catani:1996vz} which we have
to take into account order by order in $1/\mtop$, of course. The result
is a \nlo{} Monte Carlo program for $H+$parton production in gluon
fusion which, in addition to the already available pure heavy-top
limit\cite{deFlorian:1999zd,Glosser:2002gm,Ravindran:2002dc}, also
includes the first formally subleading term in $1/\mtop^2$.

We have performed a number of checks on our results. The two most
important ones are the numerical comparison of the leading terms in
$1/\mtop$ with the non-resummed part of the program {\tt
  HqT}~\cite{Bozzi:2003jy,Bozzi:2005wk,deFlorian:2011xf} where we find
agreement at the sub-percent level. The amplitudes for the
$1/\mtop^2$-terms have been checked previously by the agreement of the
inclusive cross section between Ref.\cite{Harlander:2009mq} and
\cite{Pak:2009dg}. Their proper implementation into a Monte Carlo
program is checked by the independence of the numerical results on the
so-called $\alpha$-parameter~\cite{Nagy:2003tz,Nagy:1998bb} which allows
to restrict the phase space of the dipole terms.

\subsection{Notation}

We introduce the following notation for the individual terms in our expansions:
\begin{equation}
\begin{split}
[\dd\sigma^{(l)}_k]_{ij}^\text{\abbrev X}\,,\qquad
\text{\abbrev X}\in\{\lo{},\nlo{}\}\,,\qquad
i,j\in \{q,\bar q,g\}\,,
\end{split}
\end{equation}
where $l$ denotes the order of perturbation theory, $k$ the order of the
expansion in $1/\mtop$, and {\abbrev X} the order of the \pdf{}s and the
running of $\alpha_s$ that have been used\footnote{We use the central
  {\abbrev MSTW2008} \pdf{} sets; related uncertainties are not the
  subject of this paper. See Refs.\cite{Alekhin:2010dd,Thorne:2011kq},
  however.}.  The subscript $ij$ denotes the particular partonic
channel that was taken into account. If any of the indices $l$, $k$ or
$ij$ are absent, it means that these indices are summed over all
possible values. For example,
\begin{equation}
\begin{split}
\dd\sigma^\lo{} \equiv [\dd\sigma^{(0)}]^\lo\,,\quad
\dd\sigma^\nlo{} \equiv [\dd\sigma^{(0)} +
\dd\sigma^{(1)}]^\nlo\,,
\end{split}
\end{equation}
are the \lo{} and the \nlo{} differential cross sections with exact
$\mtop$-dependence, and summed over all parton channels (recall, however,
that we neglect all $q\bar q$ and $qq$ contributions in this paper).

In order to isolate the individual corrections to the cross section, we
define the quantities
\begin{equation}
\begin{split}
[R^{(l)}_k(b)]_{ij} &=
\frac{[\dd\sigma^{(l)}_k(b)]^\nlo_{ij}}{\dd\sigma^\lo(b)}\,,\\
[K_n(b)]_{ij}&=\frac{\sum_{k=0}^n[\dd \sigma_k^{(0)}(b)+
  \dd\sigma_k^{(1)}(b)]_{ij}^\nlo}{%
  \sum_{k=0}^n[\dd \sigma_k^{(0)}(b)]_{ij}^\lo}\,.
\label{eq::randk}
\end{split}
\end{equation}
On the right hand side of these definitions, it is understood that $\dd
\sigma(b)$ is integrated over all kinematical variables {\it except} the
set $b$, where we consider $b=\{ \pt{}\}$, $b=\{y\}$, and $b=\emptyset$
(i.e., transverse momentum and rapidity distributions, and the
integrated cross section with $\pt{}>30$\,GeV). Also, if $ij$ is to be
summed over, this applies separately to the numerator and the
denominator in $[K_n(b)]_{ij}$.  The ratio $R^{(l)}_k(b)$ allows for a direct
comparison of the perturbative (index $l$) and the mass effects (index
$k$). The quantity $K(b)$, on the other hand, shows the influence of the
mass terms on the perturbative correction factor. For example, $K_0$ is
the \nlo{} K-factor in the heavy-top limit which -- in the case of the
total inclusive cross section -- has been found to approximate the exact
\nlo{} K-factor extremely well. Using the $1/\mtop{}$ expansion, we will
study whether this observation can be expected to carry over also to
differential quantities.

\subsection{Inclusive Higgs plus jet production}

\begin{figure}
  \begin{center}
    \begin{tabular}{c}
      \mbox{\includegraphics[height=.4\textheight]{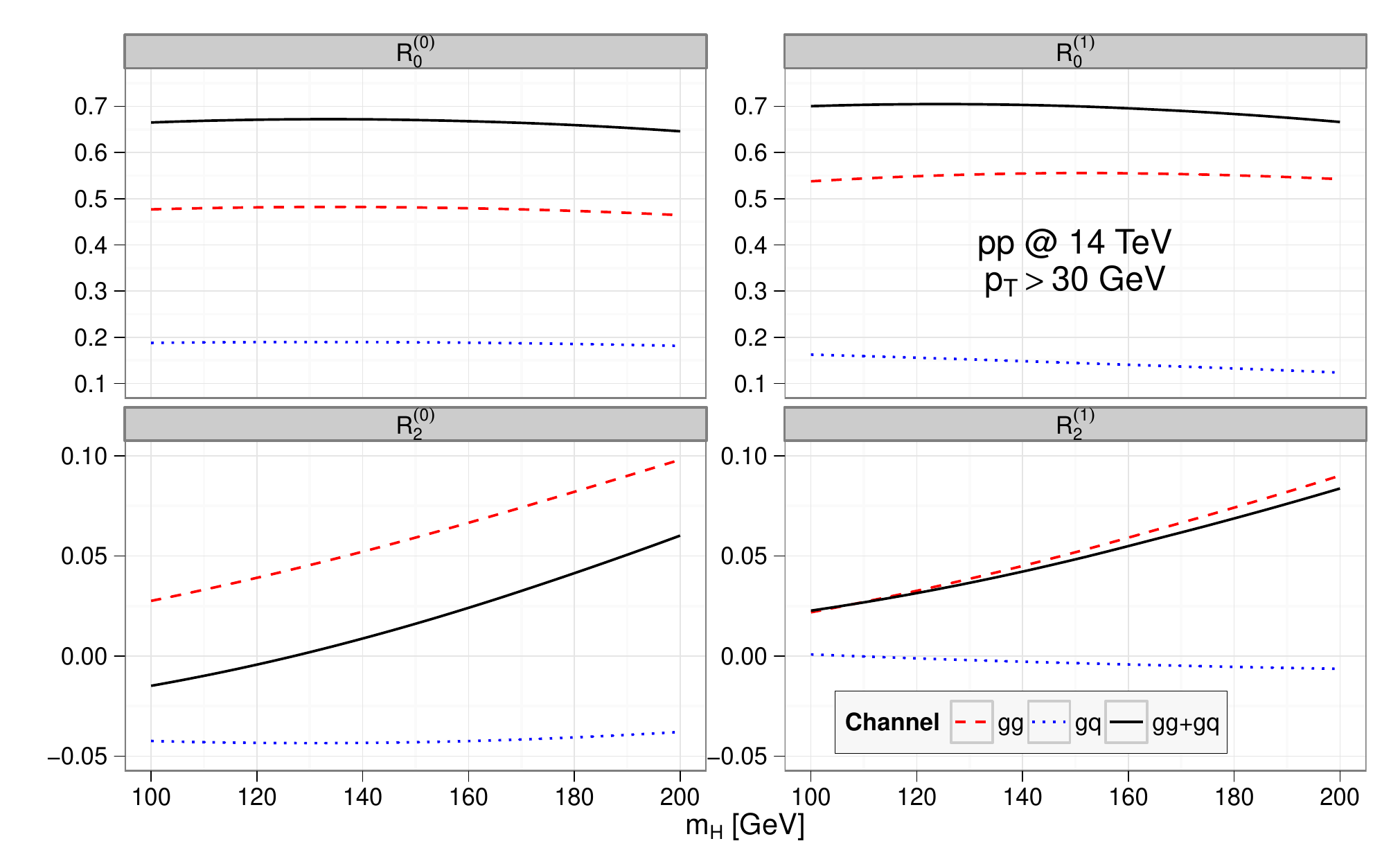}}
    \end{tabular}
    \parbox{.9\textwidth}{
      \caption[]{\label{fig::mHNLO}\sloppy Relative size of the
        perturbative and the mass effects on the integrated cross
        section, $R_k^{(l)} \equiv R_k^{(l)}(\emptyset)$, see
        \eqn{eq::dsdpt} and \eqn{eq::randk}. Upper row: \nlo{} effects
        arising solely from the \pdf{}s and the running of $\alpha_s$
        (left), and from the \nlo{} perturbative coefficient in the
        cross section (right).  Lower row: mass effects from the \lo{}
        (left) and the \nlo{} (right) perturbative coefficient. Dotted:
        only $qg$; dashed: only $gg$; solid: sum of $gg$ and $qg$.}}
  \end{center}
\end{figure}

The first observable we study is the integrated cross section for
Higgs+jet production, defined in \eqn{eq::dsdpt}.  \fig{fig::mHNLO}
compares the \nlo{} perturbative corrections to the mass effects at
\lo{} and at \nlo{}, split into the two numerically dominant
sub-channels $gg$ (dotted) and $qg$ (dashed), as well as for the sum of
both channels (solid).  The size of $R_0^{(0)} (\equiv
R_0^{(0)}(\emptyset))$ is mostly determined by the reduced value of
$\alpha_s$ when going from \lo{} to \nlo{} parton
densities. $R_0^{(1)}$, on the other hand, reflects the well-known
largeness of the perturbative effects to the gluon fusion cross
section. Considering the fact that $R_0^{(1)}$ includes a factor
$\alpha_s/\pi$ relative to $R_0^{(0)}$, it is remarkable that they are
both almost equally large. Note also that both $R_0^{(0)}$ and
$R_0^{(1)}$ depend only very weakly on the Higgs mass $\mhiggs$.

The same feature holds for the mass effects, shown in the lower two
plots, separately for the \lo{} (left) and the \nlo{} (right)
coefficients.  Note also that there is a cancellation between the $gg$
and the $qg$ channels, although much less pronounced at \nlo{} than at
\lo{}. The overall mass effects between $\mhiggs=100$\,GeV and 200\,GeV
range from $-2\%$ to 6\% for the \lo{}, and from 2\% to 8\% for the \nlo{}
coefficient. They are thus much smaller than the perturbative effects.
As expected, the mass effects decrease for smaller Higgs masses.

Concerning the K-factor, for the total inclusive cross section it has
been found to depend only very weakly on the top quark
mass~\cite{Spira:1995rr,Harlander:2009mq,Pak:2009dg}. The product of the
K-factor with the exact \lo{} cross section is thus an excellent
approximation of the higher order cross section.

\begin{figure}
  \begin{center}
    \begin{tabular}{c}
      \mbox{\includegraphics[height=.4\textheight]{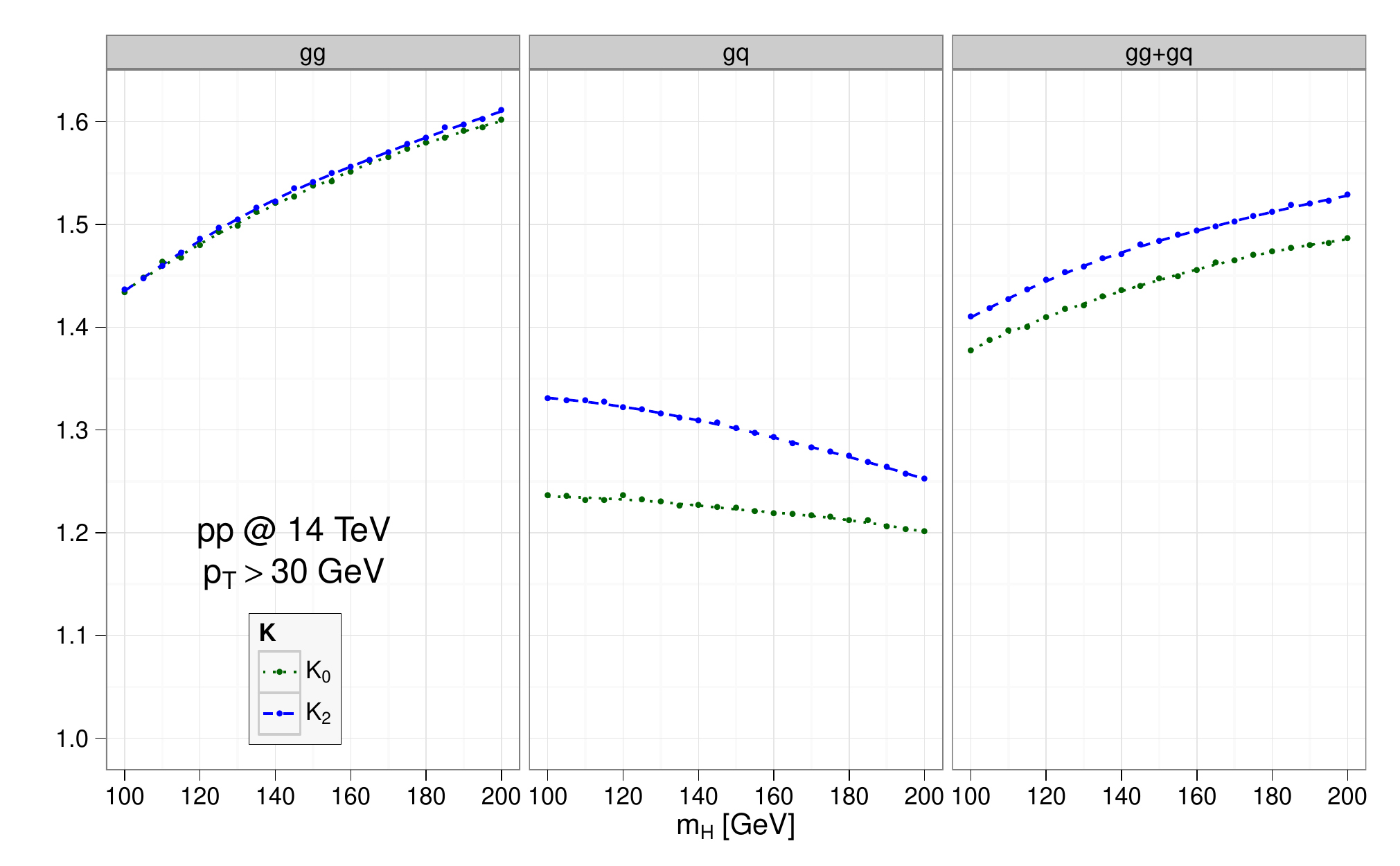}}
    \end{tabular}
    \parbox{.9\textwidth}{
      \caption[]{\label{fig::Kincl}\sloppy K-factors as defined in
        \eqn{eq::randk}, for the integrated cross section, i.e.\,
        $K_n\equiv K_n(\emptyset)$. Left/center/right plot:
        $[K_n]_{gg}/[K_n]_{qg}/[K_n]_{gg+qg}$. Dotted/dashed: $n=0/2$.
        The dots show the results of our calculation; the lines have
        been introduced to guide the eye. The deviation between the dots
    and the lines indicates our numerical error.}}
  \end{center}
\end{figure}

Let us study the extent to which we can draw a similar conclusion for
the cross section with a lower $\pt{}$ cut.  In \fig{fig::Kincl}, we
compare the result for the K-factor including mass terms, $K_2$ ($\equiv
K_2(\emptyset)$, cf.\,\eqn{eq::randk}) to the pure heavy-top
limit. Again, we consider separately the channels $gg$ and $qg$, as well
as their sum. Note however, that according to the definition in
\eqn{eq::randk}, $[K_n]_{ij}$ really only refers to the $ij$ channel,
both in the numerator {\it and} the denominator. Therefore,
$K_n \neq [K_n]_{gg}+[K_n]_{qg}$.

The agreement between $K_0$ and $K_2$ for the $gg$ channel is truly
remarkable; for the $qg$ channel, we find 5-10\% difference, but due to
the numerical dominance of $gg$, the overall agreement between $K_0$ and
$K_2$ is around 3\%.

\subsection{Transverse momentum distribution}\label{sec::pt}

\begin{figure}
  \begin{center}
    \begin{tabular}{ccc}
      \mbox{\includegraphics[height=.4\textheight]{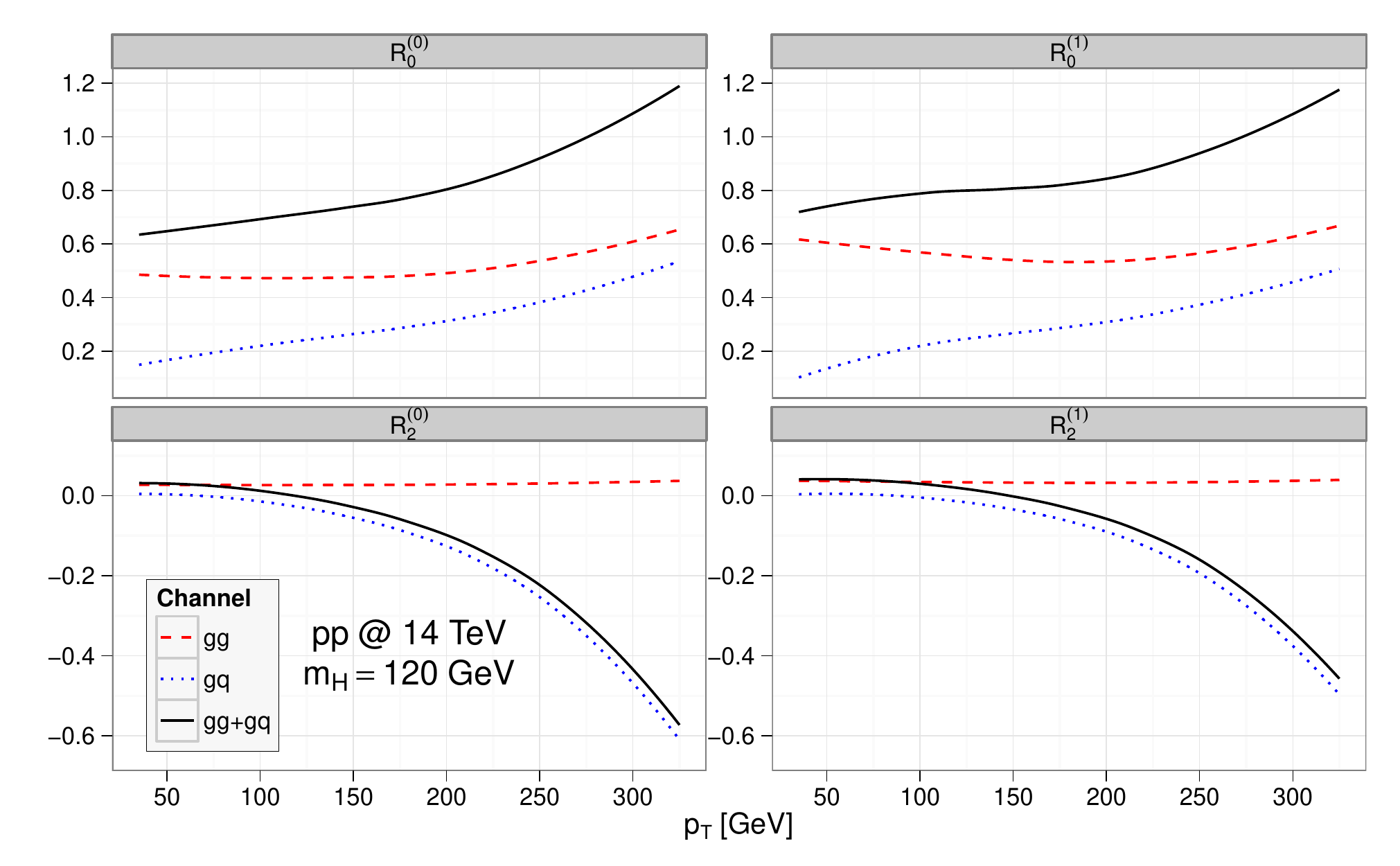}}
    \end{tabular}
    \parbox{.9\textwidth}{
      \caption[]{\label{fig::rpt}\sloppy Similar to \fig{fig::mHNLO},
        but for the differential cross section $\dd\sigma/\dd\pt{}$;
        here, $R_k^{(l)}\equiv R_k^{(l)}(\{\pt{}\})$.
    }}
  \end{center}
\end{figure}

\fig{fig::rpt} shows the $\pt{}$-dependent ratio $R(\pt{})$ ($\equiv
R(\{\pt{}\})$, cf.~\eqn{eq::randk}), in analogy to \fig{fig::mHNLO}.
The qualitative features of the individual corrections for this
differential quantity are very similar to the integrated ones.  An
observation that deserves to be pointed out is the stunning similarity
of the plots for the mass effects at \lo{} and \nlo{}.

Both at \lo{} and \nlo{}, the mass terms in the $gg$-channel are very
small ($\sim 2\%$) and almost independent of $\pt{}$, even up to
$\pt{}=300$\,GeV. The $qg$-channel behaves worse, but its mass terms
still do not amount to more than 6\% below $\pt{}=150$\,GeV; at \lo{},
they reach up to almost 50\% at $\pt{}=300$\,GeV though. The \nlo{} mass
terms are only slightly smaller. The sum of $gg$ and $qg$, however,
remains below 35\% at \nlo{} for $\pt{}<300$\,GeV; below
$\pt{}=150$\,GeV, they amount to not more than 3\%.  The mass effects
reduce the absolute value of the cross section in the $qg$-channel
significantly for $\pt{}>150$\,GeV, which is also visible at \lo{} in
\fig{fig::pTLO}, while it just slightly affects the sum of both
channels.

\begin{figure}
  \begin{center}
    \begin{tabular}{c}
      \mbox{\includegraphics[height=.4\textheight]{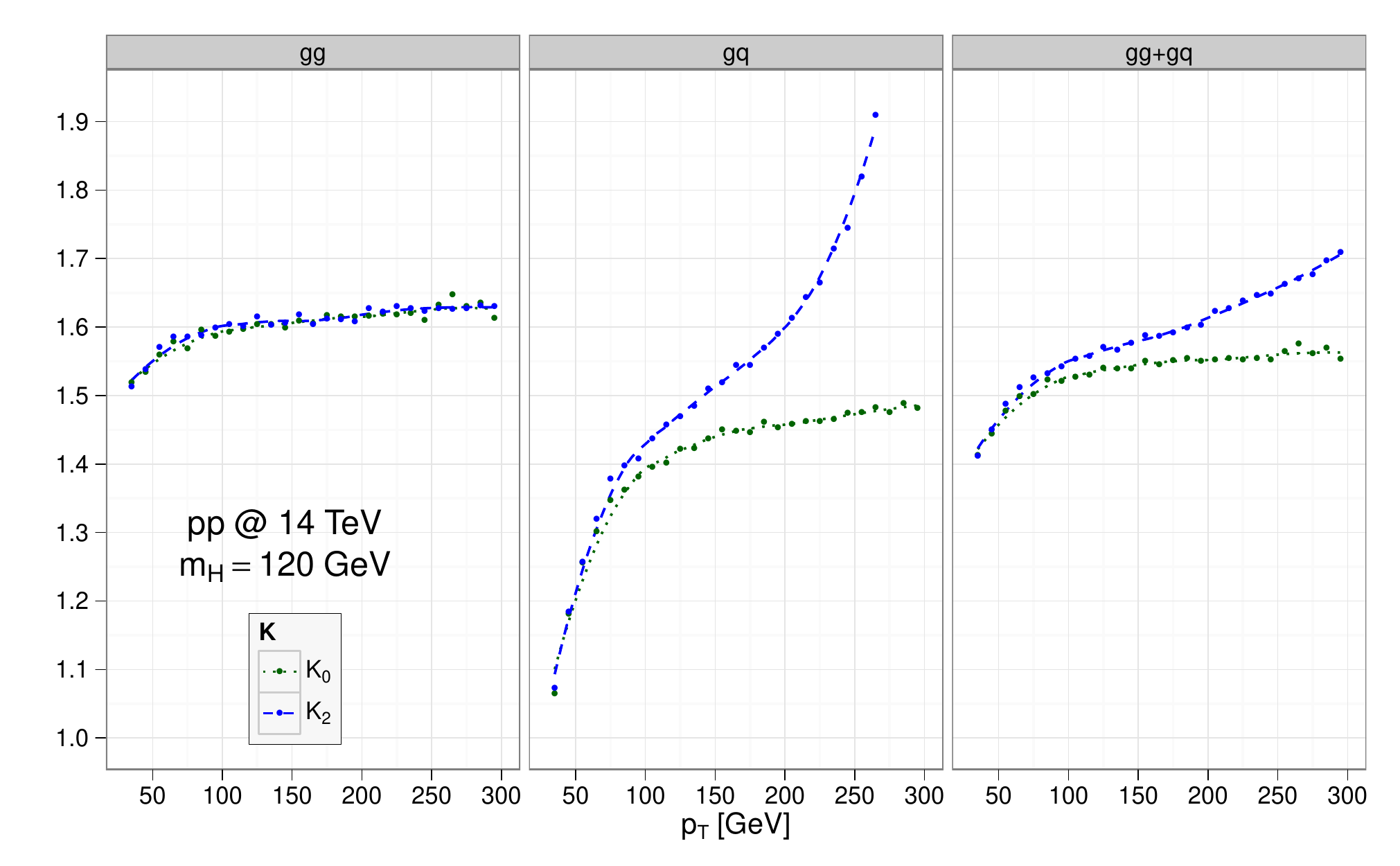}}
    \end{tabular}
    \parbox{.9\textwidth}{
      \caption[]{\label{fig::kpt}\sloppy 
        Similar to \fig{fig::Kincl},
        but for the differential cross section $\dd\sigma/\dd\pt{}$;
        here, $K_n\equiv K_n(\{\pt{}\})$.}}
  \end{center}
\end{figure}

The $\pt{}$-dependent K-factors $K_n(\pt{})$ ($\equiv K_n(\{\pt\})$) are
shown in \fig{fig::kpt}. The K-factors including leading and
subleading mass terms, $K_0$ and $K_2$, are almost identical in the
$gg$-channel. For the $qg$-channel, on the other hand, the \qcd{}
corrections to the subleading mass terms behave very differently to
the leading terms in $1/\mtop$ once $\pt{}>150$\,GeV. In the sum of
both channels, the difference remains below 3\% for $\pt{}<150$\,GeV,
and reaches 10\% at $\pt{}=300$\,GeV.

In conclusion, the behaviour of $K_2$ with respect to $K_0$ suggests
that, also for the $\pt{}$-distribution, the \qcd{} {\it corrections}
can be safely calculated in the heavy-top limit; the accuracy remains
within 2\% (10\%) below $\pt{}=150$\,GeV ($\pt{}=300$\,GeV). The
absolute distributions, however, should be calculated at \lo{} using the
full top-mass dependence, and then reweighted by these \qcd{}
corrections.

\subsection{Rapidity distribution}

\begin{figure}
  \begin{center}
    \begin{tabular}{ccc}
      \mbox{\includegraphics[height=.4\textheight]{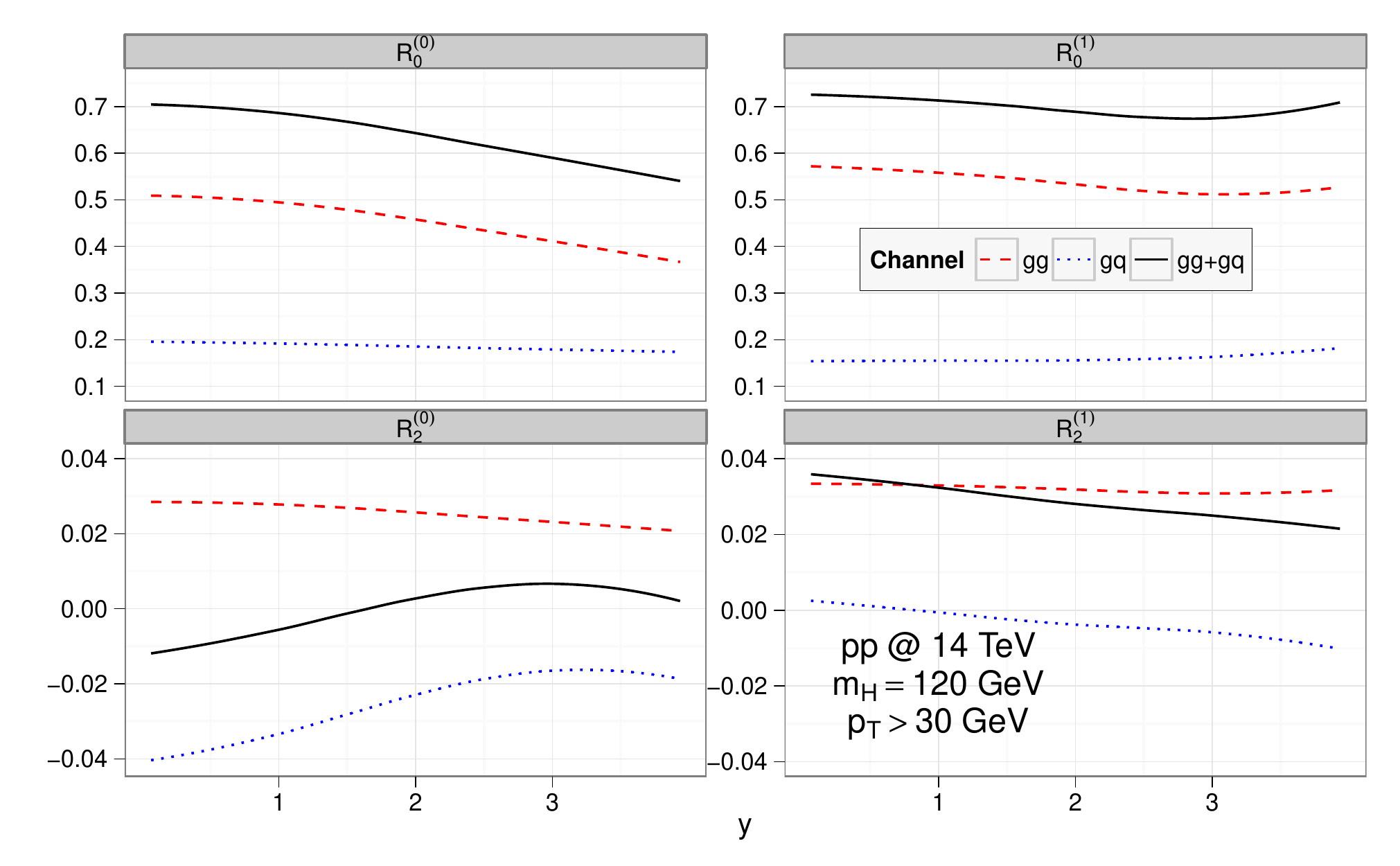}}
    \end{tabular}
    \parbox{.9\textwidth}{
      \caption[]{\label{fig::ry}\sloppy
        Similar to \fig{fig::mHNLO},
        but for the differential cross section $\dd\sigma/\dd y$;
        here, $R_k^{(l)}\equiv R_k^{(l)}(\{y\})$.}}
  \end{center}
\end{figure}

\fig{fig::ry} shows the $y$-dependent ratios $R(y) \equiv R(\{y\})$. The
perturbative effects, $R_0^{(0)}$ and $R_0^{(1)}$, are qualitatively
very similar to the quantities discussed before.  The mass effects are
generally very small over the full $y$-range. At \lo{}, there is a
significant cancellation between the $gg$- and the $qg$- channel, so
that the sum remains below 1\% almost everywhere, even though the
individual channels reach up to 4\%. At \nlo{}, the mass effects in the
$qg$-channel are very small, but due to the numerical dominance of the
purely gluon induced contributions, the overall effect reaches up to
4\%.

\begin{figure}
  \begin{center}
    \begin{tabular}{c}
      \mbox{\includegraphics[height=.4\textheight]{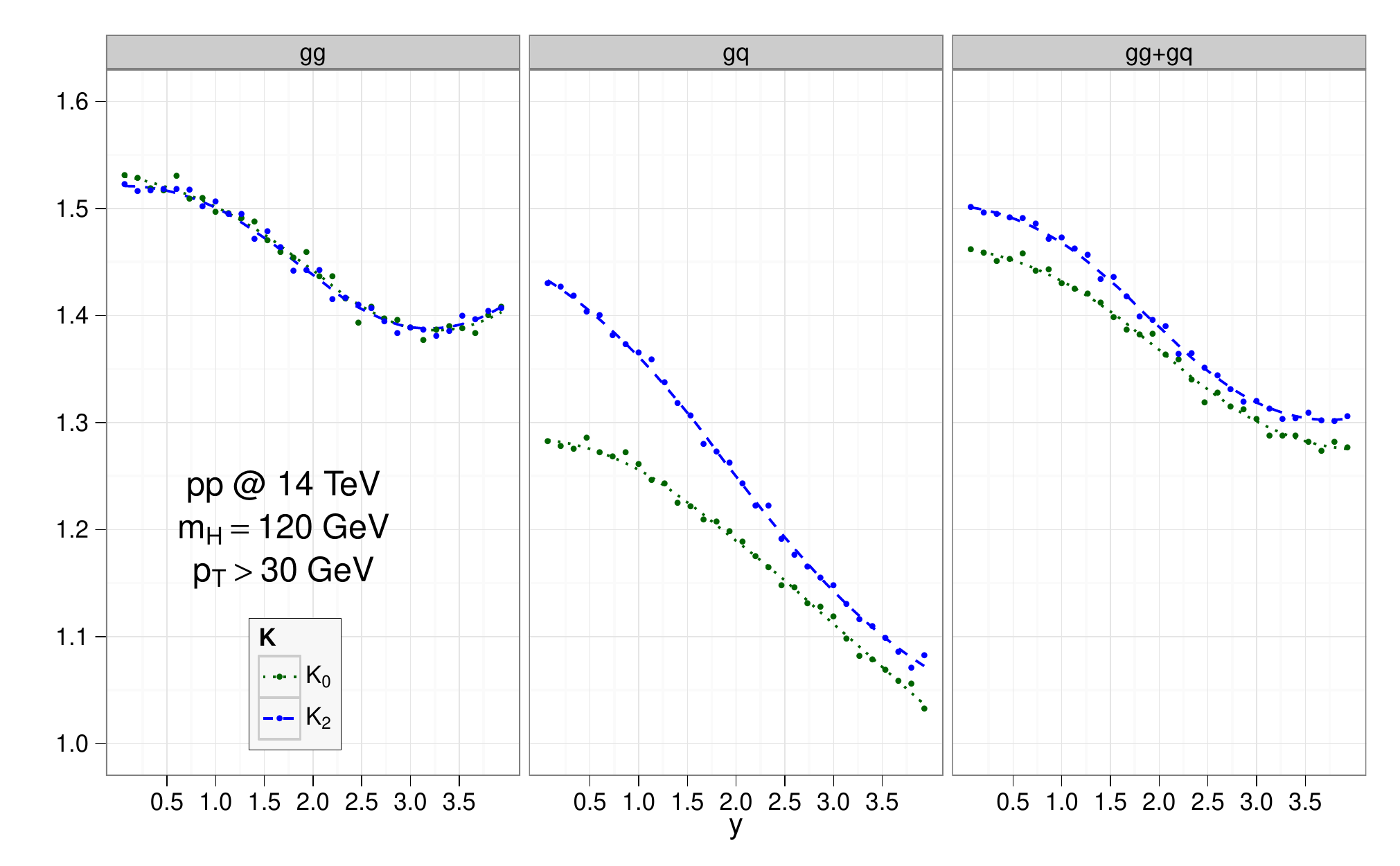}}
    \end{tabular}
    \parbox{.9\textwidth}{
      \caption[]{\label{fig::ky}\sloppy 
        Similar to \fig{fig::Kincl},
        but for the differential cross section $\dd\sigma/\dd y$;
        here, $K_n\equiv K_n(\{y\})$.}}
  \end{center}
\end{figure}

The K-factors, shown in \fig{fig::ky}, display a similar behavior as
in the previous observables: the corrections in the $gg$-channel are
practically the same in the $1/\mtop^0$- and the $1/\mtop^2$-terms, the
$qg$-channel shows some difference, but in the sum of both channels, the
$K_2$ is approximated by $K_0$ to within about 3\%. Apparently, the bad
convergence of the $qg$-channel for $\pt{} > 150$\,GeV as observed in
section \ref{sec::pt} affects $\pt{}$-integrated quantities only at the
percent level, see also \fig{fig::Kincl}.

\section{Conclusions}\label{sec::conclusions}

In this paper, the quality of the heavy-top limit in the gluon fusion
cross section has been studied.  Subleading terms in $1/\mtop{}$ have
been calculated for the $H+$jet cross section at \nlo{} \qcd{}, and
their effects on the Higgs' transverse momentum and rapidity
distribution have been evaluated.

We found that, similar to the leading terms in $1/\mtop{}$, the
perturbative corrections on the subleading terms are of order one. In
fact, the perturbative effects on the mass corrections in the
$gg$-channel are remarkably similar to those on the leading mass terms,
both for the $\pt{}$- and the $y$-distribution, as well as for the
integrated cross section of \eqn{eq::dsdpt}. The \nlo{} K-factors with
and without mass terms are therefore almost identical for this channel
alone. Including the $qg$-channel spoils this similarity to some extent,
but we still claim that the procedure of correcting the full \lo{}
prediction (including top mass effect) by the K-factor as evaluated in
the heavy-top limit provides an excellent approximation to the full
\nlo{} result, valid at the 2-3\% level for $\pt{}<150$\,GeV and for
$\pt{}$-integrated quantities. We have checked that this result holds
for Higgs masses below $2\mtop$. The accuracy is thus better than the
current uncertainty on the cross section due to its dependence on the
\pdf{}s and due to missing higher order \qcd{} corrections.

\paragraph{Acknowledgements.}
This work was supported by {\abbrev BMBF} contract 05H09PXE
and {\abbrev LHCP}heno{\abbrev N}et PITN-GA-2010-264564. 
The work of KO was supported by the US Department of Energy under
contract DE–FG03–91ER40662.

\def\app#1#2#3{{\it Act.~Phys.~Pol.~}\jref{\bf B #1}{#2}{#3}}
\def\apa#1#2#3{{\it Act.~Phys.~Austr.~}\jref{\bf#1}{#2}{#3}}
\def\annphys#1#2#3{{\it Ann.~Phys.~}\jref{\bf #1}{#2}{#3}}
\def\cmp#1#2#3{{\it Comm.~Math.~Phys.~}\jref{\bf #1}{#2}{#3}}
\def\cpc#1#2#3{{\it Comp.~Phys.~Commun.~}\jref{\bf #1}{#2}{#3}}
\def\epjc#1#2#3{{\it Eur.\ Phys.\ J.\ }\jref{\bf C #1}{#2}{#3}}
\def\fortp#1#2#3{{\it Fortschr.~Phys.~}\jref{\bf#1}{#2}{#3}}
\def\ijmpc#1#2#3{{\it Int.~J.~Mod.~Phys.~}\jref{\bf C #1}{#2}{#3}}
\def\ijmpa#1#2#3{{\it Int.~J.~Mod.~Phys.~}\jref{\bf A #1}{#2}{#3}}
\def\jcp#1#2#3{{\it J.~Comp.~Phys.~}\jref{\bf #1}{#2}{#3}}
\def\jetp#1#2#3{{\it JETP~Lett.~}\jref{\bf #1}{#2}{#3}}
\def\jphysg#1#2#3{{\small\it J.~Phys.~G~}\jref{\bf #1}{#2}{#3}}
\def\jhep#1#2#3{{\small\it JHEP~}\jref{\bf #1}{#2}{#3}}
\def\mpl#1#2#3{{\it Mod.~Phys.~Lett.~}\jref{\bf A #1}{#2}{#3}}
\def\nima#1#2#3{{\it Nucl.~Inst.~Meth.~}\jref{\bf A #1}{#2}{#3}}
\def\npb#1#2#3{{\it Nucl.~Phys.~}\jref{\bf B #1}{#2}{#3}}
\def\nca#1#2#3{{\it Nuovo~Cim.~}\jref{\bf #1A}{#2}{#3}}
\def\plb#1#2#3{{\it Phys.~Lett.~}\jref{\bf B #1}{#2}{#3}}
\def\prc#1#2#3{{\it Phys.~Reports }\jref{\bf #1}{#2}{#3}}
\def\prd#1#2#3{{\it Phys.~Rev.~}\jref{\bf D #1}{#2}{#3}}
\def\pR#1#2#3{{\it Phys.~Rev.~}\jref{\bf #1}{#2}{#3}}
\def\prl#1#2#3{{\it Phys.~Rev.~Lett.~}\jref{\bf #1}{#2}{#3}}
\def\pr#1#2#3{{\it Phys.~Reports }\jref{\bf #1}{#2}{#3}}
\def\ptp#1#2#3{{\it Prog.~Theor.~Phys.~}\jref{\bf #1}{#2}{#3}}
\def\ppnp#1#2#3{{\it Prog.~Part.~Nucl.~Phys.~}\jref{\bf #1}{#2}{#3}}
\def\rmp#1#2#3{{\it Rev.~Mod.~Phys.~}\jref{\bf #1}{#2}{#3}}
\def\sovnp#1#2#3{{\it Sov.~J.~Nucl.~Phys.~}\jref{\bf #1}{#2}{#3}}
\def\sovus#1#2#3{{\it Sov.~Phys.~Usp.~}\jref{\bf #1}{#2}{#3}}
\def\tmf#1#2#3{{\it Teor.~Mat.~Fiz.~}\jref{\bf #1}{#2}{#3}}
\def\tmp#1#2#3{{\it Theor.~Math.~Phys.~}\jref{\bf #1}{#2}{#3}}
\def\yadfiz#1#2#3{{\it Yad.~Fiz.~}\jref{\bf #1}{#2}{#3}}
\def\zpc#1#2#3{{\it Z.~Phys.~}\jref{\bf C #1}{#2}{#3}}
\def\ibid#1#2#3{{ibid.~}\jref{\bf #1}{#2}{#3}}
\def\otherjournal#1#2#3#4{{\it #1}\jref{\bf #2}{#3}{#4}}
\newcommand{\jref}[3]{{\bf #1} (#2) #3}
\newcommand{\hepph}[1]{\href{http://arxiv.org/abs/hep-ph/#1}{\tt [hep-ph/#1]}}
\newcommand{\arxiv}[2]{\href{http://arxiv.org/abs/#1}{\tt [arXiv:#1]}}
\newcommand{\bibentry}[4]{#1, {\it #2}, #3\ifthenelse{\equal{#4}{}}{}{, }#4.}

\end{document}